\documentclass{nature}
\usepackage{graphicx}
\usepackage{color}

\usepackage[table]{xcolor}
\usepackage{ctable}

\def\ket#1{|#1\rangle}
\def\one{{\mathchoice {\rm 1\mskip-4mu l} {\rm 1\mskip-4mu l} {\rm
\mskip-4.5mu l} {\rm 1\mskip-5mu l}}}

\title{Experimental magic state distillation for fault-tolerant quantum computing}

\author{Alexandre M. Souza$^1$, Jingfu Zhang$^1$, Colm A. Ryan$^1$ \& Raymond Laflamme$^{1,2}$}

\begin{document}

\maketitle

\definecolor{c1}{rgb}{0.6,0.6,0.7}
\definecolor{c2}{rgb}{0.9,0.9,1.0}
\definecolor{c3}{rgb}{1.00,0.95,0.95}
\definecolor{c4}{rgb}{0.9,1.0,0.9}

\newcommand{\mm}[1]{\Large $\mathbf{#1}$}

\newcommand{\corA}{\cellcolor{c1}}
\newcommand{\corB}{\cellcolor{c2}}
\newcommand{\corC}{\cellcolor{c3}}
\newcommand{\corD}{\cellcolor{c4}}
\begin{affiliations}
 \item Institute for Quantum Computing and Department of
Physics and Astronomy, University of Waterloo, Waterloo, Canada,
ON N2L 3G1.
 \item Perimeter Institute for Theoretical
Physics, Waterloo, Canada, ON N2J 2W9.
\end{affiliations}

\begin{abstract}
 Any physical quantum device for quantum information
processing is subject to errors in implementation. In order to be
reliable and efficient, quantum computers will need error
correcting or error avoiding methods.  Fault-tolerance achieved
through quantum error correction
 will be an integral part of quantum computers.  Of
 the many methods that have been discovered to implement it,
 a highly successful approach has been to use transversal
  gates and specific initial states.  A critical element
   for its implementation is the availability of high-fidelity
    initial states such as $\ket{0}$ and the {\em Magic State}.
Here we report an experiment, performed in a nuclear magnetic
resonance (NMR) quantum processor, showing sufficient quantum
control
     to improve the fidelity of imperfect initial magic states by
     distilling five of them into one with higher fidelity.
\end{abstract}
 Quantum information processing (QIP)
\cite{nielsen,preskill,laflamme,preskill2}  promises a dramatic
computational speed-up over classical computers for certain
problems. In implementation, the physical quantum devices for
QIP are subject to errors due to the
effects of unwanted interactions with the environment or quantum
control imperfections. In order to be reliable and efficient,
quantum computers will need error correcting or error avoiding
methods. One method to achieve fault-tolerant quantum computation
is to encode the state of a single quantum bit (qubit) into blocks
of several qubits that are more robust to errors. Based on this
idea, quantum error correction codes, the theory of fault-tolerant
quantum computation and the accuracy threshold theorem have been
developed \cite{knill1998, Kitaev1997, Aharonov}. A key element
for fault-tolerant  quantum computation is to avoid bad error
propagation. One straightforward protocol is to use transversal
gates where an error occurring on the $k$th qubit in one block can
only propagate to the $k$th qubit in the other blocks. A highly
successful approach to achieve fault-tolerant universal quantum
computation is based on quantum error correcting codes with gates
from the Clifford group that can be applied transversally
\cite{knillN04,trans1}. Unfortunately they are not universal
\cite{trans2,zengPRA78012353} and they must be supplemented with
the preparation of not only the $\ket{0}$ state but also another
type of state such as a {\em Magic State}
\cite{kitaev,magic101,magic102,magic103,magic104}. Thus, a
critical element for fault-tolerance is the availability of
high-fidelity magic states. Consequently, in the pursuit of
experimental fault-tolerant quantum computation, it is important
to determine whether we have sufficient experimental control to
prepare these magic states. In general these will be prepared with
some imprecision. The states can be improved by distilling many
magic states to produce a fewer number of them which  have higher
fidelity.  Here we report an experiment, performed in a
seven-qubit nuclear magnetic resonance (NMR) quantum processor,
showing sufficient quantum control to implement a distillation
protocol
 based on the five-bit quantum error correcting code
\cite{kitaev,5error} which uses only Clifford gates. The fidelity
of imperfect initial magic states is improved by distilling five
of them into one with higher fidelity.

\section*{Results}
{\bf Theoretical protocol.} The Clifford group is defined as the
group of operators that maps the Pauli group onto itself under
conjugation. The Pauli group is defined as \cite{nielsen}
$\{\pm\one, \pm i\one, \pm\sigma_x, \pm i\sigma_x, \pm\sigma_y,
\\ \pm i\sigma_y, \pm\sigma_z, \pm i\sigma_z\}$ where $\sigma_x$,
$\sigma_y$, $\sigma_z$, and $\one$ denote the Pauli matrices and
identity operator, respectively. The Clifford group on $n$ qubits
is a finite subgroup of the unitary group $U(2^n)$ and can be
generated by the Hadamard gate $H$, the phase-shift gate $S_{ph}$,
and the
controlled-not gate $CNOT$ 
represented as
\begin{equation} H =
\frac{1}{\sqrt{2}}\left( \begin{array}{cc} 1 & 1  \\ 1 & -1
\end{array} \right),
S_{ph} = \left( \begin{array}{cc} 1 & 0  \\ 0 & i \end{array}
\right), CNOT = \left( \begin{array}{cc} \one & 0  \\ 0 & \sigma_x
\end{array} \right) \label{clifford}
\end{equation}
in the computational basis $\{|0 \rangle, |1 \rangle\}$.

An arbitrary one-qubit state can be represented in the Bloch
sphere as
\begin{equation}
\rho =  \left ( \one + p_x \sigma_x + p_y \sigma_y + p_z \sigma_z
\right )/2 \label{anyrohgene}
\end{equation}
where $p_x$, $p_y$ and $p_z$ are the three polarization components
of the state. The magic states \cite{kitaev} are defined as the
$8$ states with $p_x = \pm 1/\sqrt{3}$, $p_y = \pm 1/\sqrt{3}$,
$p_z = \pm 1/\sqrt{3}$ ($T$ type) and the $12$ states with $p_x =
0$, $p_y = \pm 1/\sqrt{2}$, $p_z = \pm 1/\sqrt{2}$; $p_y = 0$,
$p_z = \pm 1/\sqrt{2}$, $p_x = \pm 1/\sqrt{2}$; $p_z = 0$, $p_x =
\pm 1/\sqrt{2}$, $p_y = \pm 1/\sqrt{2}$ ($H$ type).
These states are called ''magic'' because of their ability, with
Clifford gates, to enable universal quantum computation and the
ability to be purified, when it has been prepared imperfectly,
using only Clifford group operations \cite{kitaev}. In our current
work we distill an imperfect magic state into a $T$-type magic
state represented as
\begin{equation}
\rho_M =  \left [ \one + (\sigma_x +\sigma_y + \sigma_z)/\sqrt{3}
\right ]/2. \label{rohgene}
\end{equation}
To quantify how near a state $\rho$ is to the magic state, we
define the m-polarization (polarization in the direction of the
magic state)
\begin{equation}
p =  2 \, Tr [\rho_M\rho] - 1   = \frac{1}{\sqrt{3}} \left ( p_x +
p_y + p_z \right). \label{pol}
\end{equation}

The distillation algorithm requires five copies of a faulty
magic state  $\rho_{in} = \rho^{\otimes 5}$ as the input state.
In the original proposal \cite{kitaev}, the measurement of four
stabilizers $S_i$ ($i=1$, ..., $4$) is applied to $\rho_{in}$,
where $S_1 = \sigma_x \otimes \sigma_z \otimes \sigma_z \otimes
\sigma_x \otimes \one$, $S_2 = \one \otimes \sigma_x \otimes
\sigma_z \otimes \sigma_z \otimes \sigma_x$, $S_3 = \sigma_x
\otimes \one \otimes \sigma_x \otimes \sigma_z \otimes \sigma_z$
and $S_4 = \sigma_z \otimes \sigma_x \otimes  \one \otimes
\sigma_x \otimes \sigma_z$. If the outcome of any of these
observables is $-1$, the state is discarded and the distillation
fails. If the results of all the measurement are $+1$,
corresponding to the trivial syndrome, one applies the decoding
transformation for the five-qubit error correcting code
\cite{5error} to the measured state and obtains the output state
$\rho_{dis} \otimes |0000 \rangle \langle 0000|$ where
$\rho_{dis}$ has the output m-polarization $p_{out}$. If
 the input m-polarization $p_{in} > p_0= \sqrt{3/7} \approx
0.655$, distillation is possible and $p_{out}>p_{in}$ and produces
a state nearer to the magic one. In an iterative manner, it is
possible to obtain the output m-polarization approaching $1$.

As NMR QIP is implemented in an
ensemble of spin systems, only the output of expectation
values of ensemble measurements \cite{Jones} are available. Consequently the
above projective measurement of the stabilizers cannot be
implemented in our experiment. However, as the decoding operation
is just a basis transformation from one stabilizer subspace to
another, it is possible to evaluate the result of the distillation
after decoding. Therefore, we directly apply the decoding
operation to the input state $\rho_{in}$, and the output state
becomes a statistical mixture of 16 possible outcomes represented
as
\begin{eqnarray}\label{dists}
\rho_{out} =\sum_{i=0}^{15}\theta_{i}\rho_{i} \otimes  |i \rangle
\langle i |
\end{eqnarray}
where $\theta_i$ is the probability of each outcome, and
$|i\rangle=|0000\rangle$, $|0001\rangle$, ... , $|1111\rangle$,
for $i=0$, $1$,  $2$, ..., $15$,
noting $\rho_0=\rho_{dis}$. Now measuring $\ket{0}$ on all four
qubits in $\ket{i}$ indicates a successful purification.  We can
obtain $\theta_i$ and $\rho_{i}$ using partial quantum state
tomography \cite{tomo}.

{\bf Experimental results.} The data were taken with a Bruker 700
MHz spectrometer. We choose $^{13}$C- labelled trans-crotonic acid
dissolved in d6-acetone as a seven-qubit register.
The structure of the molecule  and the parameters 
of qubits are shown in Table \ref{mol}.
We prepare a labelled pseudo-pure state
$\rho_{s}=\mathbf{0}\mathbf{0}\sigma_z
\mathbf{0}\mathbf{0}\mathbf{0}\mathbf{0}$  using the method in
Ref. \cite{knill}, where 
$\mathbf{0}=|0\rangle\langle0|$ and the order of qubits is M,
H$_1$, H$_2$,  C$_1$, C$_2$, C$_3$, C$_4$. One should note that we
are using the deviation density matrix formalism.



We prepare an initial imperfect magic state with three equal
polarization components by depolarizing the state $\mathbf{0} =
(\one+\sigma_z)/2$. First we apply a $\pi/2$ pulse to rotate the
state $\mathbf{0}$ to $(\one+\sigma_{x})/2$ and then another
$\pi/2$ pulse along direction $[\cos a,\sin a,0]$ is applied. We
use phase cycling to average
the $x$- and $y$- components of the state to zero, and therefore
the polarization of the spin initially in the state $\mathbf{0}$
is reduced. The depolarized state is represented as
\begin{equation}\label{oneDepo}
(\one -\sigma_{z}\sin a)/2.
\end{equation}
Finally we apply a rotation with angle $\arccos(1/\sqrt{3})$ about the
direction $[-1/\sqrt{2}, 1/\sqrt{2}, 0]$ to
obtain an imperfect magic state
\begin{equation}\label{rohgene1}
\rho =  \left [ \one + p(\sigma_x +  \sigma_y + \sigma_z)/\sqrt{3}
\right ]/2
\end{equation}
where $p=-\sin a$. The evolution of $\sigma_z$ in the preparing
$\rho$ is shown in Fig. \ref{bloch}.  By doing the above operation
for qubits M, C$_1$, C$_2$, C$_3$, C$_4$, respectively, we obtain
five copies of the imperfect magic states $\rho_{in} = \rho
^{\otimes 5}$. Exploiting partial state tomography, we measure $p$
for each qubit and use the average as the input m-polarization
$p_{in}$ for $\rho_{in}$.

The circuit for the distillation operation is shown in Fig.
\ref{dist}.
 C$_1$ carries the distilled state after the
distillation.
With partial state tomography, we can determine 
$\theta_i$ 
and  $\rho_i$ in equation (\ref{dists}), where
$\rho_{0}=\rho_{dis}$, from which the output m-polarization
$p_{out}$ is obtained. The experimental results for magic state
distillation for various $p_{in}$  are shown in Fig.
\ref{poldist},
 where
figures (a) and (b) show the measured $p_{out}$ and $\theta_{0}$,
respectively. The straight line in figure (a) represents the
function $p_{out}=p_{in}$.  The data points above the line show
the states that have been distilled experimentally.

The implementation time of the distillation procedure is about 0.1
s, a non-negligible amount of time
(10\%) compared to coherence time ($T_{2}$ in Table \ref{mol}). 
Hence the decay of the signals due to the limitation of coherence
time is an important source of errors. We extract $p_{out}$ by
measuring the ratio of $\theta_{0} p_{out}$ and $\theta_{0}$,
where these two factors are obtained by various single coherent
terms in a series of experiments (see details in Methods). We have
assumed that the terms have the same amount of decoherence. The
results of simulations with dephasing rates $T_2^{*}$ and $T_2$
are shown in Fig. \ref{poldist}
 as the blue squares and red triangles. 
The simulation results show that the decoherence rates are long
enough to allow the distillation and suggest that the deviation of
$\theta_0$ from the theoretical expectation can be mainly
attributed to relaxation effects. Additionally, imperfection in
the shaped pulses and inhomogeneities of magnetic fields
also contribute to errors.

\section*{Discussion}
We modify the original distillation protocol by avoiding the
projective measurement, which is not possible to implement in the
NMR QIP's. We exploit partial-state tomography to obtain each output in the mixture of the outcomes
after the distillation, and only in a post-processing step do we
choose the one we need. Although we could access the $|0000\rangle\langle0000|$ subspace
 using a procedure similar to the pseudo-pure state
preparation, the method would take a substantially longer amount of
time and would be more error-prone. In this work we aim for a
quantitative result, i.e. increasing the magic state purity. We
need to minimize the readout manipulations to avoid control-error
induced distortions of the inferred final state and associated
purity.  Hence, we limited ourselves to simple high-fidelity
readout procedures.

In summary, we have implemented a protocol for distilling magic
states based on the five qubit quantum error correction code. We
exploit five qubits by controlling a seven-qubit NMR quantum
information processor. The experiment shows that we have obtained
enough control to purify faulty magic states through distillation.

\begin{methods}
{\bf Overview of the experiment.} To implement the experiment, we
exploit standard Isech and Hermite-shaped pulses as well as
numerically optimized GRAPE pulses \cite{kaneja} to implement
single-spin operations. The GRAPE pulses are optimized to be
robust to radio frequency (r.f.) inhomogeneities and chemical
shift variations. All pulses are combined in a custom-built
software compiler \cite{pra}. The compiler loads the information
about the internal Hamiltonian and the desired unitary
transformation from simple predefined building blocks. The blocks
are then systematically put together to form a pulse sequence
ensuring that the errors in the building blocks do not propagate
as the sequence progresses.

{\bf R.f. selection.} The effect of pulse imperfections due to
r.f. inhomogeneities is reduced by selecting signal based on r.f.
power \cite{knill}. The signal selection is achieved by spatially
selecting molecules from a small region in the sample. The method
is similar to imaging methods \cite{image} and has been used in
previous works \cite{knill}. Here we substitute the original pulse
sequence proposed in Ref.\cite{knill} by a single GRAPE pulse to
optimize the performance. Besides reducing r.f inhomogeneities,
the spatial selection of spins can also reduce the static field
inhomogeneities and therefore reduces the loss of signal during
the experiment. We have found that the effective relaxation time
($T_2^*$) of spins after the r.f. selection increases
significantly, e.g., up to a factor 2 for some spins.

{\bf Partial state tomography.} We use the spectra obtained from
the labelled pseudo-pure state
$\rho_{s}=\mathbf{0}\mathbf{0}\sigma_z
\mathbf{0}\mathbf{0}\mathbf{0}\mathbf{0}$ shown in Fig.
\ref{ppsfig}
  as a phase reference  and to normalize the signals in C$_1$ and C$_2$
spectra,  for measuring the initial and output
m-polarization. To obtain the reduced density matrix of C$_1$
through the partial state tomography, we expand equation
(\ref{dists}) as a sum of product operators \cite{tomo}, and
represent  $\rho_{i}$ as
\begin{equation}
 \rho_{i} = \frac{1}{2} \left ( I + p_{i,x} \sigma_x + p_{i,y} \sigma_y + p_{i,z} \sigma_z \right ). \label{rohgene}
\end{equation}
In the expansion there are 128 terms that are required to
determine by the experiment.

  The coefficients of such expansion can be directly related to the
 measurable spectral amplitudes \cite{tomo}. On the other hand, such
coefficients can also be related to the relevant parameters of
(\ref{dists}), i.e, $p_{i,x}$, $p_{i,y}$, $p_{i,z}$ and $\theta_i$
for $i=0$, $1$, $2$, ..., $15$. The relation between these
parameters and the NMR observables can be expressed by the set of
linear equations
\begin{equation}
C = A \times R. \label{seteq}
\end{equation}
The $nth$ element, $C(n)$, of the column vector $C$ is the
coefficient related to the operator $\sigma_z^{\bar{n}_4}I\sigma_z
s \sigma_z^{\bar{n}_1}\sigma_z^{\bar{n}_2}\sigma_z^{\bar{n}_3}$
with the order of qubits $M$, H$_1$, H$_2$,  C$_1$, C$_2$, C$_3$,
C$_4$,  where $s$ can be one element of the Pauli group
$\{\sigma_x, \sigma_y, \sigma_z, I\}$ and the vector $\bar{n} =
(\bar{n}_1, \bar{n}_2,  \bar{n}_3,  \bar{n}_4) $ is the four digit
binary representation of the integer $n-1$ .
For $s = \sigma_x$, $\sigma_y$, $\sigma_z$, and $I$, $R(n) =
\theta_n p_{n,x}$, $\theta_n p_{n,y}$, $\theta_n p_{n,z}$ and
$\theta_n$, respectively. The elements of the matrix $A$ are given
by
\begin{equation}
A(k,m) = \Pi_{i=1}^{4} (-1)^{\bar{k}_i  \bar{m}_i }.
\end{equation}

Providing that we have all necessary coefficients measured, we can
reconstruct the distilled states using the following approach.
First, we fit the NMR spectral lines to the yield complex
amplitudes for measuring all necessary coefficients \cite{tomo}.
Fig. \ref{spectra} illustrates the spectra of C$_1$ after the
completion of distillation for $p_{in} = 0.95$, where the
experimentally measured, fitted, and ideal spectra are shown as
the red, black, and blue curves, respectively. Then the state
(\ref{dists}) is reconstructed by solving the set of equations
(\ref{seteq}). Our calculation shows that four readout operations
are sufficient to determine all coefficients: first, read out on
C$_1$; second, read out on C$_1$ after the application of a $\pi /
2$ pulse; third, read out on C$_2$ after the application of a $\pi
/ 2$ pulse; and forth, read out on C$_2$ after a polarization
transfer from H$_1$ to C$_2$. The last two readout operations are
sufficient to measure all $\theta_i$, and the first two are used
to measure $\theta_i \rho_i$. The errors for the coefficients, as
well errors for $p_{i,x}$, $p_{i,y}$, $p_{i,z}$ and $\theta_i$ ,
are estimated from the uncertainty of the fitting parameters. The
measured initial and output m-polarization, as well as $\theta_i$
and $\rho_i$, are listed as Supplementary Tables S1-S10 in the
Supplementary Information \cite{SI}. The comparison of the various
measured $\rho_0$ with the theory is shown as equations (1-7) in
the Supplementary Methods in the Supplementary Information.

\end{methods}



\begin{addendum}
 \item [Acknowledgments] We thank the Premier Discovery Award from the Government of
Ontario, Industry Canada and the Canadian Institute for Advanced
Research for financial support.  We thank Nathan
Babcock, Josh Slater, Stephanie Simmons and Martin Laforest for
discussions and their contributions in the first stage of this
experiment, and also thank Camille Negrevergne and Adam Hubbard
for laying the groundwork for the experimental work in realization
of distillation.  J.-F.Z thanks Bei Zeng for helpful
discussion.

 \item[Competing Interests] The authors declare that they have no
competing financial interests.
 \item[Correspondence] Correspondence and requests for materials should be
addressed to
J.-F.Z. (j87zhang@iqc.ca). \item [Author contributions] A.M.S. and
J.-F.Z. performed the experiment and  the numerical simulations,
C. A.R. helped design the experimental scheme and wrote the
control software; R.L. conceived the ideas and supervised the
experiment. All authors contributed to the writing of the paper,
discussed the experimental procedures and results.

\newpage
\begin{table}  
\centering \addtolength{\tabcolsep}{-5pt}
  \begin{tabular}{cccccccc}
\bottomrule
 \corA       & \corA \mm{M} & \corA \mm{H_1}&  \corA \mm{H_2}&  \corA \mm{C_1}&  \corA \mm{C_2}&  \corA \mm{C_3}&  \corA \mm{C_4} \\
 \corA \mm{M}   & \corB \mm{-1309} & & & & & &  \\
 \corA \mm{H_1} &\corC \mm{6.9} &\corB \mm{-4864}& && & &  \\
 \corA \mm{H_2} &\corC \mm{-1.7} &\corC \mm{15.5}& \corB \mm{-4086}& & & &  \\
 \corA \mm{C_1} &\corC\mm{ 127.5} &\corC \mm{3.8}&\corC  \mm{6.2}&\corB  \mm{-2990}& & &  \\
 \corA \mm{C_2} &\corC\mm{ -7.1} &\corC \mm{156.0}&\corC  \mm{-0.7}& \corC \mm{41.6}& \corB \mm{-25488}& &  \\
 \corA \mm{C_3} &\corC \mm{6.6} &\corC \mm{-1.8}& \corC \mm{162.9}& \corC \mm{1.6}&\corC  \mm{69.7}& \corB \mm{-21586}&  \\
 \corA \mm{C_4} &\corC \mm{-0.9} &\corC \mm{6.5}& \corC \mm{3.3}& \corC \mm{7.1}&\corC  \mm{1.4}& \corC \mm{72.4}&\corB  \mm{-29398} \\  \toprule
 \corD \mm{T_2(s)}&\corD \mm{0.84} &\corD \mm{0.85}& \corD \mm{0.84}& \corD \mm{1.27}&\corD  \mm{1.17}&\corD  \mm{1.19}& \corD \mm{1.13} \\
 \corD \mm{T^*_2(s)}& \corD \mm{0.61} &\corD \mm{0.57}&\corD \mm{0.66}&\corD \mm{1.04}&\corD \mm{0.66}&\corD \mm{1.16}& \corD \mm{0.84} \\
 \end{tabular}
  \caption {{\bf Characteristics of the molecule of trans-crotonic
acid.} Molecular structure is shown as Fig. \ref{crot7}. The
chemical shifts and J-coupling constants (in Hz) are on and below
the diagonal in the table, respectively. The transversal
relaxation times $T_2$ measured by a Hahn echo and $T_2^{*}$
calculated by measuring the width of the peaks through fitting the
spectra are listed at the bottom. The chemical shifts are given
with respect to reference frequencies of 700.13 MHz (protons) and
176.05 MHz (carbons). The molecule contains nine weakly coupled
spin half nuclei but consists of a seven qubit system since the
methyl group can be treated as a single qubit using a
gradient-based subspace selection \cite{knill}.}  \label{mol}
\end{table}


\newpage
\begin{figure}[tbp]
\begin{center}
\includegraphics[width=9cm]{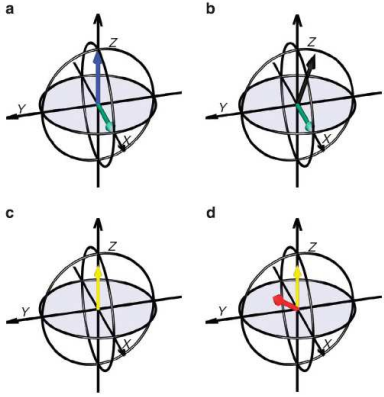}
\end{center}
\caption {{\bf The evolution of $\sigma_z$ in the preparation of a
faulty magic state in Bloch sphere, where arrows represent the
states of the qubit.} {\bf a}, A $\pi/2$ rotation along y-axis
transforms $\sigma_z$ (blue) to $\sigma_x$ (green). {\bf b},
Another $\pi/2$ rotation along direction $[\cos a, \sin a, 0]$
transforms $\sigma_x$ (green) to $\sigma_x\cos^{2}a +\sigma_y\cos
a \sin a -\sigma_z\sin a $ (black). In phase cycling we apply the
second $\pi/2$ rotation by changing $a$ to $\pi +a$ to transform
$\sigma_x$ to $\sigma_x\cos^{2}a +\sigma_y\cos a \sin a
+\sigma_z\sin a$. After averaging the $x$- and $y$- components to
zero, the polarization is reduced to $-\sigma_z \sin a$, shown as
the yellow arrow in figure {\bf c}, noting that $a\in [\pi,
3\pi/2]$. {\bf d}, A final rotation with angle $\arccos
(1/\sqrt{3})$ along $[-1/\sqrt{2}, 1/\sqrt{2}, 0]$ transforms
$-\sigma_z \sin a$ (yellow) to $ -\sin a(\sigma_x +  \sigma_y +
\sigma_z)/\sqrt{3}$ (red).}\label{bloch}
\end{figure}

\newpage
\begin{figure}[tbp]
\begin{center}
\includegraphics[width=14cm]{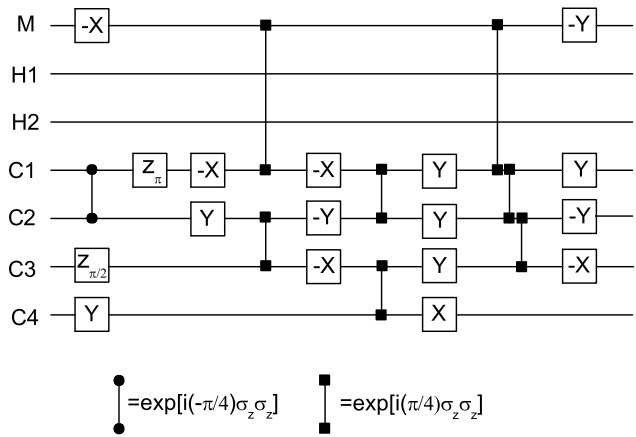}
\end{center}
\caption{{\bf Gate sequence for Magic State distillation.} The
sequence is constructed from the five qubit error correction code
\cite{5error} where $\pm X=\exp(\mp i \pi \sigma_{x}/4)$, $\pm
Y=\exp(\mp i \pi \sigma_{y}/4)$, and $Z_{\alpha}=\exp(- i \alpha
\sigma_{z}/2)$. Qubits labelled as M, C$_1$, C$_2$, C$_3$ and
C$_4$ are used to encode the five copies of the initial state. Due
to the nature of the algorithm, the carbon C$_1$ contains the
distilled magic state only when  M, C$_2$, C$_3$ and C$_4$ are in
the $|0000 \rangle$ state. It is important to emphasize that all
gates are Clifford gates. The refocussing pulses (which also
decouple $H_1$ and $H_2$) are not shown.} \label{dist}
\end{figure}

\newpage
\begin{figure}[tbp]
\begin{center}
\includegraphics[width=12cm]{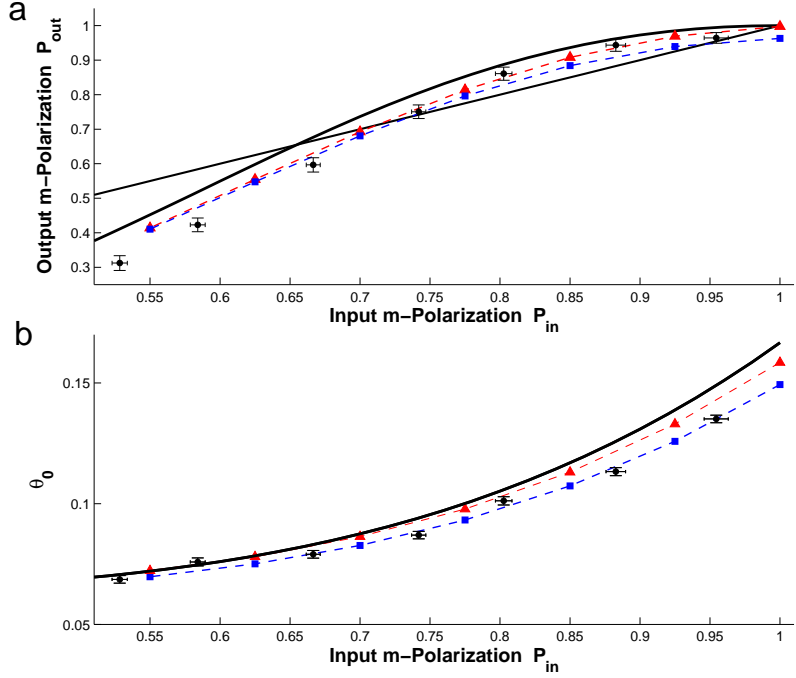} 
\end{center}
\caption{{\bf Experimental results after the completion of magic
state distillation.} Output m-polarization of the faulty magic
state ({\bf a}) and the probability $\theta_{0}$ ({\bf b}) of
finding this state in the mixture of outcomes [see equation
(\ref{dists})] as a function of the input m-polarization of the
initial faulty magic state. The experimental data are represented
by the filled circles and the error bars are estimated from the
uncertainty of the fitting parameters. The line in figure (a)
represents the function $p_{out} = p_{in}$. The experimental
points above the line show the states that have been distilled,
while the points below the line show the states that cannot be
distilled in the protocol. The theoretical prediction is
represented by the black solid curves. The blue squares and red
triangles, connected by dashed lines for visual convenience, are
the simulation results where the dephasing rates are chosen as
$T_{2}^{*}$ and $T_{2}$ (see Table \ref{mol}), respectively. The
effective $T_2$ during the experiment should be similar to the
Hahn echo $T_2$. The deviation can be attributed to other error
sources (see text). The dephasing times of H$_{1}$ and H$_{2}$
actually do not influence the results because H$_{1}$ and H$_{2}$
can be effectively assumed in $\mathbf{0}$ and $\sigma_{z}$ during
the whole experiment, respectively. } \label{poldist}
\end{figure}

\newpage
\begin{figure}[tbp]
\begin{center}
\includegraphics[width=17cm]{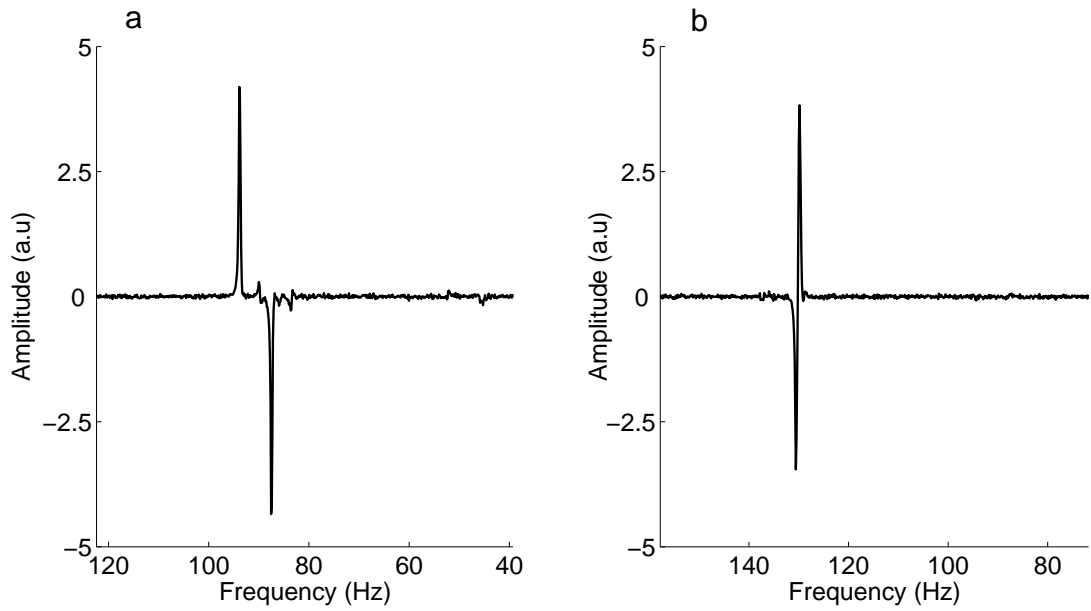} 
\end{center}
\caption{{\bf NMR spectra.} {\bf a,b,} Spectra of C$_1$ and C$_2$
obtained by $\pi/2$ readout pulses when the system lies in the
labelled pseudo-pure state $\rho_{s}=\mathbf{0}\mathbf{0}\sigma_z
\mathbf{0}\mathbf{0}\mathbf{0}\mathbf{0}$. The vertical axes have
arbitrary but the same units.}\label{ppsfig}
\end{figure}

\newpage
\begin{figure}[tbp]
\begin{center}
\includegraphics[width=17.0cm]{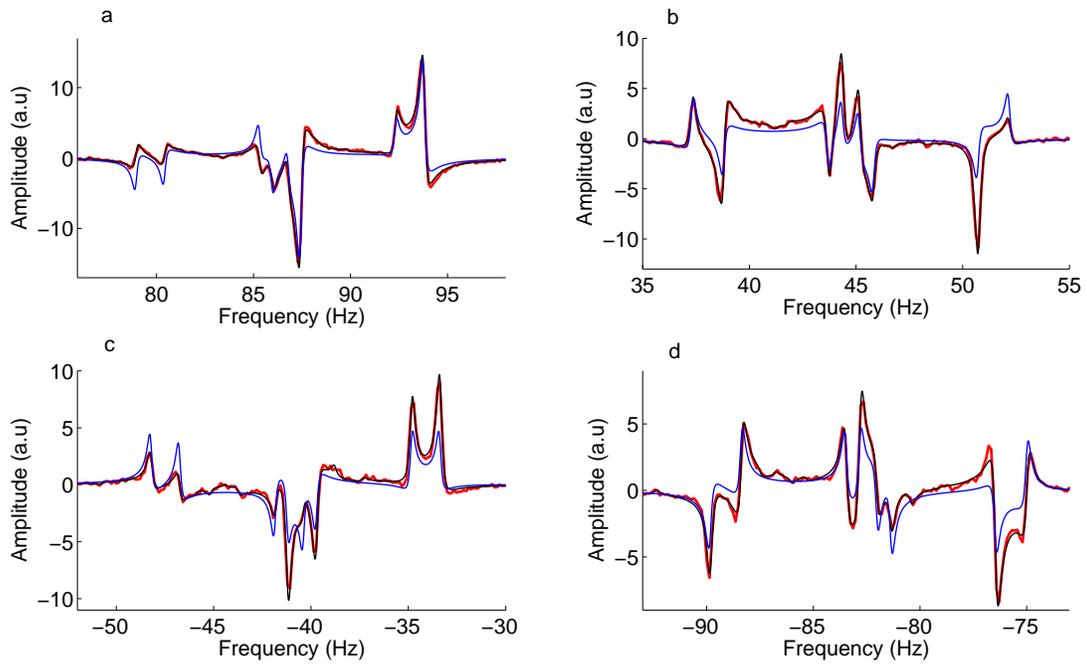} 
\end{center}
\caption{{\bf Spectra of C$_1$  after the completion of
distillation for $p_{in} = 0.95$.} The spectra are divided in four
different parts shown as figures ({\bf a-d}) for better
visualization. The vertical axes have arbitrary but the same
units. The experimentally measured, fitting, and ideal spectra are
shown as the red, black, and blue curves, respectively. }
\label{spectra}
\end{figure}

\newpage
\begin{figure}[tbp]
\begin{center}
\includegraphics[width=8cm]{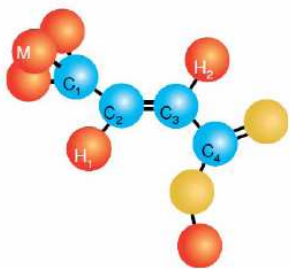} 
\end{center}
\caption{ {\bf Molecular structure of trans-crotonic acid.}}
\label{crot7}
\end{figure}

\end{addendum}


\end{document}